\documentclass[aps,prd,twocolumn,showpacs,groupedaddress]{revtex4}  
\usepackage{graphicx}  
\usepackage{dcolumn}   
\usepackage{bm}        
\usepackage{amssymb}   
\usepackage{amsmath, amsthm}

\newcommand{\ds}        {\ensuremath{{D_{s}^{-}}}}

\newcommand{\bzerod}    {\ensuremath{ B^{0}_d}}

\newcommand{\bs}        {\ensuremath{{B_{s}^{0}}}}

\newcommand{\dstokstk}    {\ensuremath{{D^-_{s} \to K^{*0} K^-}}}
\newcommand{\dstophipi}   {\ensuremath{{D^-_{s} \to \phi \pi^{-}}}}
\newcommand{\mukstk}    {\ensuremath{{\mu^{+} K^{*0} K^-}}}
\newcommand{\muphipi}   {\ensuremath{{\mu^{+} \phi \pi^-}}}
\newcommand{\ksttokpi}      {\ensuremath{{K^{*0} \rightarrow  K^+ \pi^-}}}
\newcommand{\bstodsmux} {\ensuremath{{B_{s}^{0} \to \mu^+ D_{s}^{-} X}}}

\newcommand{\phitokk}      {\ensuremath{{\phi \rightarrow  K^{+}K^{-}}}}

\begin{document}

\hspace{5.2in} \mbox{Fermilab-Pub-09-140-E}

\title{Search for CP violation in {\boldmath $\bstodsmux$} decays in
  {\boldmath $\bar{p}p$} collisions at {\boldmath $\sqrt{s}=1.96$}~TeV}

%
\author{V.M.~Abazov$^{37}$}
\author{B.~Abbott$^{75}$}
\author{M.~Abolins$^{65}$}
\author{B.S.~Acharya$^{30}$}
\author{M.~Adams$^{51}$}
\author{T.~Adams$^{49}$}
\author{E.~Aguilo$^{6}$}
\author{M.~Ahsan$^{59}$}
\author{G.D.~Alexeev$^{37}$}
\author{G.~Alkhazov$^{41}$}
% ...

\affiliation{\vspace{0.1 in}(The D\O\ Collaboration)\vspace{0.1 in}}
% ...
\affiliation{$^{6}$University of Alberta, Edmonton, Alberta, Canada;
                Simon Fraser University, Burnaby, British Columbia, Canada;
                York University, Toronto, Ontario, Canada and
                McGill University, Montreal, Quebec, Canada}
% ...
\affiliation{$^{30}$Tata Institute of Fundamental Research, Mumbai, India}
% ...
\affiliation{$^{37}$Joint Institute for Nuclear Research, Dubna, Russia}
% ...
\affiliation{$^{41}$Petersburg Nuclear Physics Institute,
                St. Petersburg, Russia}
% ...
\affiliation{$^{49}$Florida State University, Tallahassee, Florida 32306, USA}
% ...
\affiliation{$^{51}$University of Illinois at Chicago,
                Chicago, Illinois 60607, USA}
% ...
\affiliation{$^{59}$Kansas State University, Manhattan, Kansas 66506, USA}
% ...
\affiliation{$^{65}$Michigan State University,
                East Lansing, Michigan 48824, USA}
% ...
\affiliation{$^{75}$University of Oklahoma, Norman, Oklahoma 73019, USA}
% ...
\date{submitted to PRD on 20 April 2010; published 26 July 2010}

\begin{abstract}

We have performed a search for CP violation in a sample of
$\bstodsmux$ decays corresponding to $5$~fb$^{-1}$ of
proton-antiproton collisions collected by the D0 detector in Run~II at
the Fermilab Tevatron Collider. New physics in $B_s^0$ mixing could
contribute a significant CP violating weak phase, which would be
observed as a difference in the decay-time distribution for $B_s^0\to
\bar{B}_s^0$ oscillated states versus that for $\bar{B}_s^0\to
{B}_s^0$.  A fit to the decay-time distributions of the
$B_s^0/\bar{B}_s^0$ candidates yields the flavor-specific
asymmetry $a_{fs}^{s}=[-1.7\pm
9.1\mathrm{(stat)}^{+1.4}_{-1.5}\mathrm{(syst)}] \times
10^{-3}$, which excludes CP violation due to new physics within the experimental
sensitivity.

\end{abstract}

\maketitle 

\section{Introduction}

The search for large CP violating (CPV) effects in the
$B_s^0-\bar{B}_s^0$ system is of special interest since their
observation would be a direct indication of new physics.  A non-zero
CPV weak phase $\phi_s$ arises from the phase difference between the
absorptive and dispersive parts of the $B_s^0-\bar{B}_s^0$ mixing
amplitude:
$\phi_s=\arg(-M_{12}^{s}/\Gamma_{12}^{s})$~\cite{lenz_nierste}, where
$M_{12}^{s}$ and $\Gamma_{12}^{s}$ are the off-diagonal elements of
the $B_s^0-\bar{B}_s^0$ mass and decay matrices, respectively.  A
global fit to various measurements interpreted in the context of the
standard model (SM) yields the prediction
$\phi_s^{SM}=(4.2\pm1.4)\times10^{-3}$~\cite{lenz_nierste}. However,
new physics, such as the existence of a fourth generation~\cite{hou},
could contribute an additive phase $\phi_s^{NP}$ such that $\phi_s =
\phi_s^{SM}+\phi_s^{NP}$.  Recent measurements of $\phi_s^{NP}$ in
$\bs \to J/\psi \phi$ decays by the CDF~\cite{cdfjpsiphi} and
D0~\cite{d0jpsiphi} collaborations differ from zero by approximately two
standard deviations, which motivates further CP violation studies
in $B_{s}^{0}$ decays.

The CPV weak phase $\phi_s$ can be obtained from the flavor-specific
asymmetry 
\begin{equation}
a_{fs}^s=\frac {\Gamma_{\bar{B}_{s}^{0}(t)\to
    f}-\Gamma_{B_{s}^{0}(t)\to \bar{f}}}
{\Gamma_{\bar{B}_{s}^{0}(t)\to f}+\Gamma_{B_{s}^{0}(t)\to \bar{f}}}
\end{equation}
according to $a^s_{fs}=\frac{\Delta\Gamma_s}{\Delta m_s} \tan\phi_s$,
where $\Delta \Gamma_s$ and $\Delta m_s$ are the width and mass
differences, respectively, between the heavy and light eigenstates of
the mixed $B^{0}_{s}$ system.  World average values of these
quantities~\cite{hfag} yield $a^s_{fs}=(-8.4^{+5.2}_{-6.7})\times
10^{-3}$~\cite{private}.  Improved precision is needed to establish
evidence of physics beyond the SM, which predicts
$a^s_{fs}=(0.0206\pm0.0057)\times10^{-3}$~\cite{lenz_nierste}.

We present a measurement of $a^{s}_{fs}$ using $\bstodsmux$ decays
(charge conjugate states are assumed throughout) reconstructed in
proton-antiproton collisions collected by the D0 detector between
April 2002 and August 2008, corresponding to about $5$~fb$^{-1}$ of
integrated luminosity~\cite{lumi}.  This CPV study is complementary to
those using inclusive dimuon events and $\bs \to J/\psi \phi$
decays. The time-integrated inclusive dimuon analysis~\cite{d0dimu}
does not distinguish between the various $B$ hadrons and therefore
depends heavily on $\bzerod$ asymmetry results from the $B$ factories
and the determination of the $B^+/B_d^0/B_s^0/b$-baryon production
fractions. In contrast, the present measurement allows a
straightforward determination of the sample composition, due to the
partial reconstruction of the $B_s^0$ meson.  The $\bs \to J/\psi
\phi$ CPV measurements~\cite{cdfjpsiphi,d0jpsiphi} involve an analysis
of the decay product transversity angles to separate the CP-even and
CP-odd components. The present measurement does not require any
angular analysis. Furthermore, it uses all the $B_{s}^{0}$ production
and decay information available in an event, the former via
initial-state flavor tagging, when possible, and the latter via an
unbinned fit to the decay-time distribution.

\section{Event Reconstruction and Selection}

The D0 detector is described in detail elsewhere~\cite{D0}.  Charged
particles are reconstructed using the central tracking system, which
consists of a silicon microstrip tracker (SMT) and a central fiber
tracker (CFT), both located within a 2~T superconducting solenoidal
magnet.  An additional single-layer silicon microstrip detector called
Layer~0~\cite{l0} installed immediately outside the beam pipe provides
improved impact parameter resolution and vertexing efficiency.
Electrons are identified by the preshower detector and
liquid-argon/uranium calorimeter. Muons are identified by the muon
system, which consists of a layer of tracking detectors and
scintillation trigger counters in front of 1.8~T iron toroids,
followed by two similar layers after the toroids~\cite{NIM}.  The
solenoid and toroid polarities are reversed regularly, the latter
allowing a determination of the muon charge asymmetries induced by the
detector.  The 5~fb$^{-1}$ data sample used in this analysis is
divided into two subsamples; the first 1.3~fb$^{-1}$ is referred to as
Run~IIa and the remaining 3.7~fb$^{-1}$ collected after the
installation of the Layer~0 detector is referred to as Run~IIb.

Most of the sample was collected with single muon triggers.  The
reconstruction of the $\bstodsmux$ candidates is as follows.  The
tracks were required to have signals in both the CFT and SMT.
Muons were required to have measurements in at least two layers of the
muon system. The muon segment was required to be matched to a track in
the central tracking system and to have momentum
$p(\mu^+)>3.0$~GeV/$c$ and $p_T(\mu^+)>2.0$~GeV/$c$, where $p_T$ is
the momentum component transverse to the proton beam direction.

\begin{figure}
\includegraphics[width=0.48\textwidth]{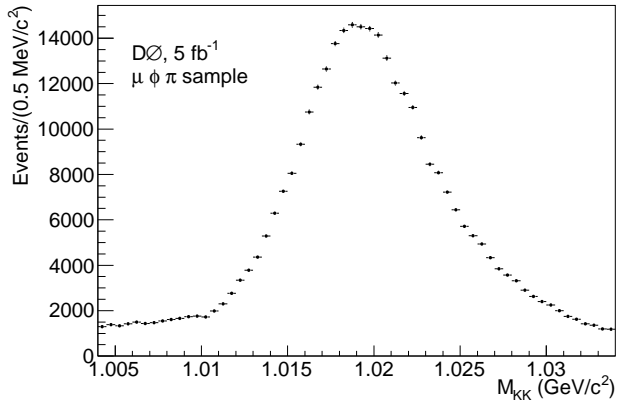}
\caption{\label{prd_fig1} 
$K^+K^-$ invariant mass distribution for the $\muphipi$ sample.  
}
\end{figure}

\begin{figure}
\includegraphics[width=0.48\textwidth]{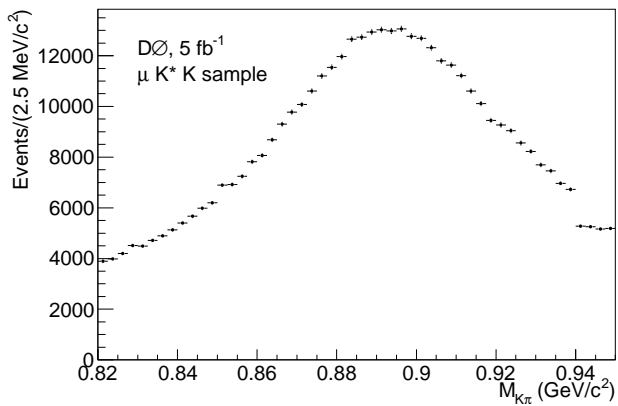}
\caption{\label{prd_fig2} $K^+\pi^-$ invariant mass distribution for
the $\mukstk$ sample. 
}
\end{figure}

All the tracks in each event were clustered into jets using the
algorithm described in Ref.~\cite{durham}. The $D_s^-$ candidates were
then formed from tracks found in the same jet as the muon candidate.
Two $\mu^+D_s^-$ final state samples were reconstructed: $\muphipi$
where $\phi \rightarrow K^{+}K^{-}$ and $\mukstk$ where $K^{*0}
\rightarrow K^{+}\pi^{-}$. The $\muphipi$ reconstruction follows the
technique described in Ref.~\cite{d0bsmix}. The $\phi$ candidate was
formed from two oppositely charged particles assigned the kaon mass
(the D0 detector is unable to distinguish between kaons, pions and
protons). The kaon candidates were required to have
$p_T(K^{\pm})>0.7$~GeV$/c$. The $K^+K^-$ invariant mass distribution
for the $\phi$ candidates in the final selected sample is shown in
Fig.~\ref{prd_fig1}.  The $\phi$ candidate was required to have an
invariant mass in the range $1.004<M(K^+K^-)<1.034$~GeV$/c^2$,
consistent with that of a $\phi$ meson. A pion candidate with charge
opposite to that of the muon and $p_T(\pi^{-})>0.5$~GeV$/c$ was then
added to form the $D_s^-$ meson candidate. In the $\mukstk$ decay
mode, the $D_s^-$ candidate was formed from three charged particles,
one with the same charge as the muon and two with a charge opposite to
that of the muon.  The particle with the same charge as the muon was
assigned the kaon mass and required to have
$p_{T}(K^{+})>0.9$~GeV$/c$.  A more stringent requirement of
$p_{T}(K^{-})>1.8$~GeV$/c$ was imposed on the second kaon candidate to
reduce combinatorial background.  The third particle was assigned the
pion mass and required to have $p_T(\pi^{-})>0.5$~GeV$/c$.  The
$K^+\pi^-$ invariant mass distribution for the $K^{*0}$ candidates in
the final selected sample is shown in Fig.~\ref{prd_fig2}.  $K^{*0}$
candidates were required to have an invariant mass in the range
$0.82<M(K^{+}\pi^{-})<0.95$~GeV$/c^{2}$, consistent with the $K^{*0}$
mass. Details about the $\mukstk$ analysis are available in
Ref.~\cite{thesis}. The sharp edge below 0.94~GeV/$c^2$ in
Fig.~\ref{prd_fig2} is an artifact of the selection criteria that are
dependent on the mass resolution. Since the events in this mass range
are primarily background, the measured asymmetry is not affected.

The primary proton-antiproton interaction vertex was determined
for each event. The average position of the collision point in the
plane transverse to the beam was measured for each run and was
included as a constraint. The precision of the primary vertex
reconstruction for each event was on average about 20~$\mu m$ in the
plane perpendicular to the beam direction and about 40~$\mu m$ along
the beam direction.

In both decay modes, a common $D_s^-$ decay vertex was formed from the
three $D_s^-$ daughter tracks using the algorithm described in
Ref.~\cite{vertex}.  To reduce combinatorial background, the $D^-_s$
vertex was required to have a displacement from the primary vertex in
the transverse plane of at least four standard deviations.  The cosine
of the angle between the $D^-_s$ momentum and the direction from the
primary vertex to the $D^-_s$ decay vertex was required to be greater
than 0.9. The trajectories of the muon and $D^-_s$ candidates were
required to be consistent with originating from a common vertex (used
as the $B^0_s$ decay vertex) and the $\mu^+ D_s^-$ system was required
to have an invariant mass between 2.6 and 5.4~GeV$/c^2$, consistent
with coming from a $B^0_s$ semileptonic decay.  We define an angle
between the combined $\mu^+D^-_s$ momentum (an approximation of the
$B^0_s$ momentum) and the direction from the primary vertex to the
$B^0_s$ decay vertex. The cosine of this angle was required to be
greater than 0.95 for $B_s^0$ candidates displaced from the primary
vertex in the transverse plane by at least four standard
deviations. These angular criteria ensure that the $D^-_s$ momentum is
sufficiently aligned with that of its $B^0_s$ parent.  The
displacement and angular criteria give rise to a decay-time dependent
reconstruction efficiency, which is discussed later.

\begin{figure}
\includegraphics[width=0.48\textwidth]{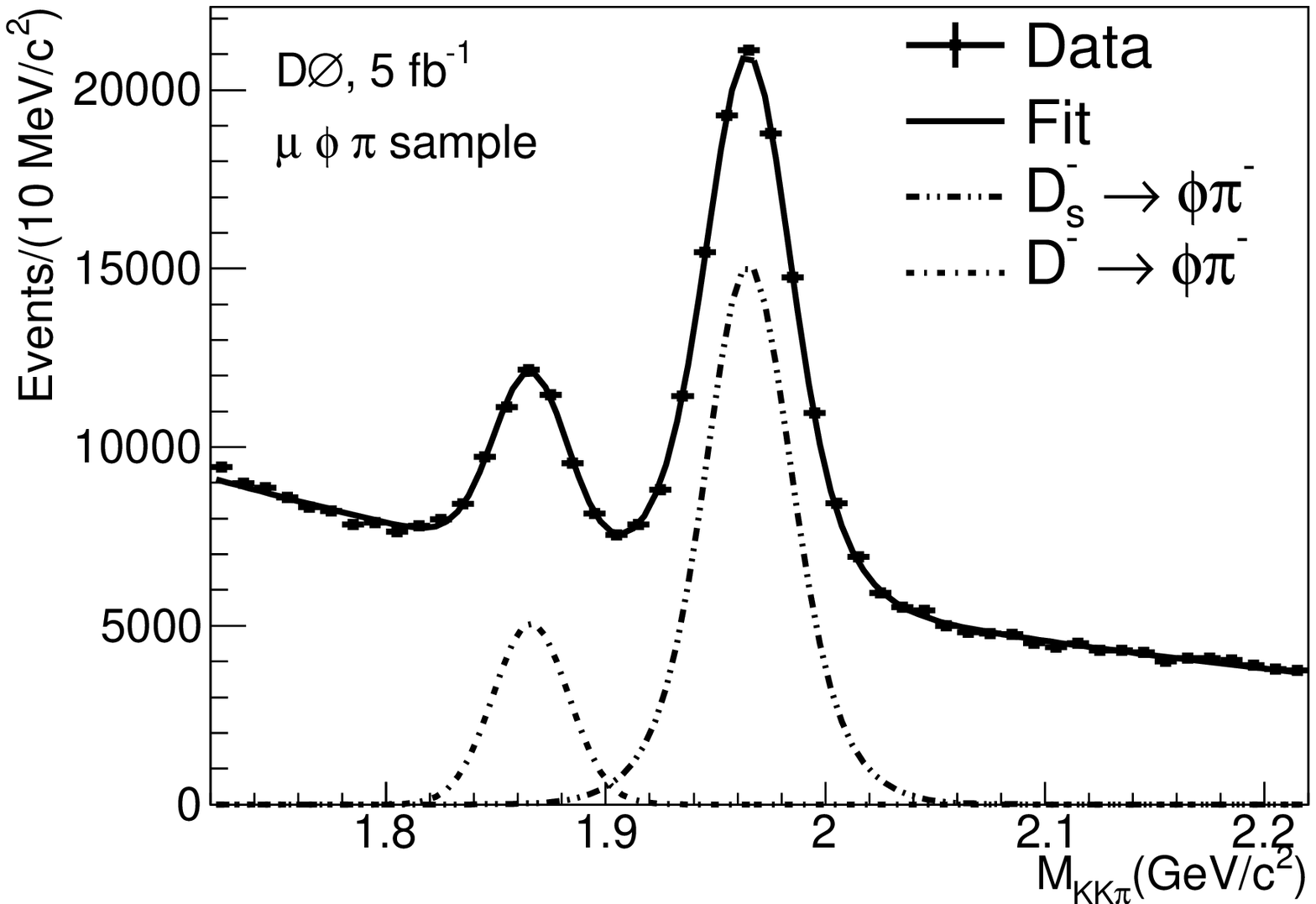}
\caption{\label{prd_fig3} 
$K^+K^-\pi^-$ invariant mass distribution for
 the $\muphipi$ sample with the solid line representing the mass fit result.  
}
\end{figure}

\begin{figure}
\includegraphics[width=0.48\textwidth]{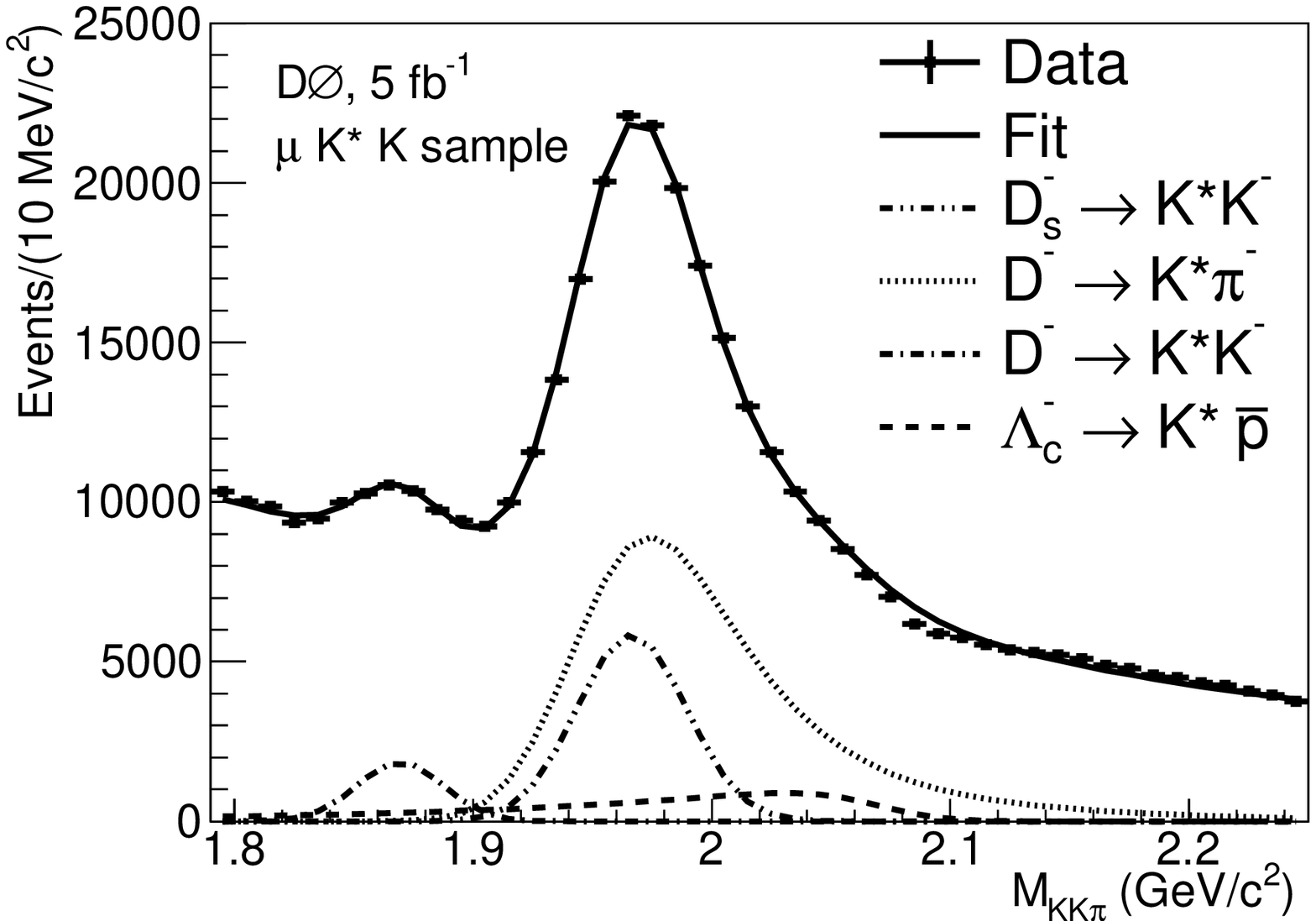}
\caption{\label{prd_fig4} 
$K^+K^-\pi^-$ invariant mass distribution for
 the $\mukstk$ sample with the solid line representing the mass fit result.  
}
\end{figure}

The $B^0_s$ selection was further improved using a likelihood ratio
method~\cite{d0bsmix,like_ratio} that combines a number of
discriminating variables: the helicity angle between the $D_s^-$ and
$K^{\pm}$ momenta in the $\phi$ or $K^{*0}$ center-of-mass frame; the
isolation of the $\mu^+ D_s^-$ system, defined as $I=p(\mu^+
D_s^-)/[p(\mu^+ D_s^-)+\Sigma p_i]$, where the sum is over all tracks
in the cone $\sqrt{(\Delta \phi)^2+(\Delta \eta)^2}<0.5$ around the
$\mu^+D_s^-$ direction ($\phi$ is the azimuthal angle of the track,
$\eta=-\ln[\tan(\theta/2)]\ $ is the pseudorapidity and $\theta$ is
the polar angle between the track momentum and the beam axis); the
$\chi^2$ of the $D_s^-$ vertex; the invariant masses $M(\mu^+ D_s^-)$,
$M(K^+ K^-)$ ($\muphipi$ sample) or $M(K^+ \pi^-)$ ($\mukstk$ sample);
and $p_T(K^+ K^-)$ ($\muphipi$ sample) or $p_T(K^-)$ ($\mukstk$
sample).  The final requirement on the likelihood ratio variable,
$y_{\mathrm{sel}}$, was chosen to maximize the predicted ratio
$S/\sqrt{S+B}$ in a data subsample corresponding to 20\% of the full
data sample, where $S$ is the number of signal events and $B$ is the
number of background events determined from signal and sideband
regions of the $M(K^+K^-\pi^-)$ distributions.  Since
$y_{\mathrm{sel}}$ is independent of the muon charge information, this
optimization does not influence the measured asymmetry and, therefore,
the subsample was also used in the complete analysis.  The numbers of
$D_s^-$ signal events, determined from a fit to the $K^+K^-\pi^-$
invariant mass distributions (see Figs.~\ref{prd_fig3}
and~\ref{prd_fig4}), are $N(\muphipi) = 81,\!394 \pm 865$ and
$N(\mukstk) = 33,\!557 \pm 1,\!200 $, where the uncertainties are
statistical.  In approximately $1\%$ of events, both a $\mukstk$
candidate and a $\muphipi$ candidate were found.  To avoid double
counting, these events were removed from the $\mukstk$ sample. The
events with two $\muphipi$ or $\mukstk$ candidates were removed from
the analysis.

\section{Flavor Tagging}

In order to measure the flavor-specific asymmetry $a_{fs}^{s}$, it is
necessary to distinguish between $B_s^0\to \bar{B}_s^0$ and
$\bar{B}_s^0\to B_s^0$ oscillated states, which requires knowledge of
the initial-state ({\em production}) and final-state ({\em decay})
flavors of the reconstructed $B_s^0$ meson.  The final-state $b$ quark
flavor is correlated with the charge of the muon in
$B^{0}_{s}\to\mu^{+}\ds X$ semileptonic decays.  The initial-state
flavor, which provides additional information in the likelihood fit
that is used to extract $a^{s}_{fs}$, was determined using an
opposite-side tagging (OST)~\cite{d0bsmix,ost} algorithm.  This
algorithm relies in most cases on the reconstruction of a second
lepton (muon or electron) from the decay of the other $b$ quark
produced in the proton-antiproton interaction.  This lepton appears on
the side of the detector opposite to the reconstructed $B_s^0$ meson
(hence the term ``opposite-side'' tag) and its charge provides the
flavor tag.  When a second lepton cannot be identified, the OST
algorithm attempts to reconstruct an opposite-side secondary vertex,
in which case the charge is determined from the tracks comprising the
vertex. Only 21\% of the events are tagged; the remaining 79\% of
events have neither a lepton nor a secondary vertex on the opposite
side. Properties of the tagging lepton and secondary vertex tracks are
incorporated in the tagging variable, $d_{\mathrm{tag}}$, which is
assigned to each $B^{0}_{s}$ candidate.  By definition, the variable
$d_{\mathrm{tag}}$ is defined in the interval \mbox{[-1,1]}. An event
with $d_{\mathrm{tag}}>0$ is tagged as an initial $b$ quark and an
event with $d_{\mathrm{tag}}<0$ is tagged as an initial $\bar{b}$
quark.  A higher magnitude $\left|d_{\mathrm{tag}}\right|$ corresponds
to a higher tagging confidence. Samples of reconstructed
$B_d^0\rightarrow \mu^+ D^{*-}X$ and $B^+\rightarrow \mu^+ D^0X$
decays were used to empirically determine the calibration function,
$\mathcal{D}(d_{\mathrm{tag}})$, called {\em dilution} (see Tables~I,
II and V in Ref.~\cite{ost}). This function is used to calculate the
probability {\sl
p}$_{\mathrm{cor}}=(\mathcal{D}(d_{\mathrm{tag}})+1)/2$ that a given
$B^0_s$ candidate has been tagged correctly.  In events where no
tagging information is available, the dilution is set to zero.

\section{Proper Decay Time}

The proper decay time of each $B^0_s$ candidate is derived from the
measured displacement $\vec{L}_T$ of the $B^0_s$ decay vertex from the
primary vertex in the transverse plane.  A Lorentz transformation of
$\vec{L}_T$ into the $B_s^0$ rest frame would yield the desired decay
time.  However, the undetected neutrino and other non-reconstructed
particles in the semileptonic $B^0_s$ decay prevent the precise
determination of $p_T(B^0_s)$ needed to calculate the Lorentz boost
factor.  Instead, the combined transverse momentum of the $\mu^+ \ds$
pair, $p_{T}(\mu^+ \ds)$, is used to calculate the visible proper
decay length (VPDL)
\begin{equation}
l = M(B^0_s) \cdot [\vec{L}_{T}\cdot \vec{p}_T(\mu^+ \ds)]/[p_{T}(\mu^+
  \ds)]^{2},
\end{equation}
where $M(B^0_s) = 5.3663$~GeV$/c^2$~\cite{pdg}. The proper decay
length of each $B^0_s$ meson is then $c t(B^0_s) = l \mathcal{K}$,
where $\mathcal{K}= p_T(\mu^+ D_s^-)/p_T(B_{s}^{0})$ is a correction
factor that accounts for the missing momentum.  Since $\mathcal{K}$ is
not known on an event-by-event basis, it was estimated from a Monte
Carlo simulation, which included the {\sc pythia}
generator~\cite{pythia} interfaced with the {\sc evtgen} decay
package~\cite{evtgen}, followed by full {\sc geant}~\cite{geant}
modeling of the detector response and event reconstruction.  As large
samples were required to obtain sufficient statistical precision, only
generator-level information was used to determine the $\mathcal{K}$
distributions.  However, a sample of fully simulated events was used
to verify that the difference between generator-level $\mathcal{K}$
distributions and those obtained using fully simulated/reconstructed
events is negligible. A model of the muon trigger efficiency
dependence on $p_T(\mu^+)$ was included in the construction of the
$\mathcal{K}$ distributions, which were obtained for each decay
channel contributing to the signal sample.  $B_s^0$ semileptonic
decays yielding an invariant $\mu^+ D_s^-$ mass that is close to the
actual $B_s^0$ mass have less missing momentum than those with lower
$M(\mu^+ D_s^-)$.  Distributions of $\mathcal{K}$ for a given decay
channel in ten bins of $M(\mu^+ D_s^-)$ were used to exploit this
fact, thereby reducing the uncertainty of the proper decay time
associated with $\mathcal{K}$.

The probability density function (PDF) for the $B^0_s$ decay time is
convoluted with the PDF describing the VPDL detector resolution and
the PDF for the $\mathcal{K}$ factor. The decay-time PDF is then
scaled by the $B^{0}_{s}$ reconstruction efficiency, which was found
for each decay channel using fully simulated events.  In the
$\muphipi$ sample, the reconstruction efficiency for the decay
channels contributing to the signal was then tuned to data by fixing
the $B^{0}_{s}$ lifetime, $\tau_{B^0_s}=1.470$~ps~\cite{pdg}, and
releasing the signal efficiency parameters in the decay-time fit to
the data.  In the $\mukstk$ sample, the overlap between $\dstokstk$
and $D^- \to K^{*0} \pi^-$ candidates (see
Fig.~\ref{prd_fig4}) prevented tuning to the data so the signal
efficiency determined from the simulation was used as is, allowing
$\tau_{B^0_s}$ to float in the fit.  For both decay modes, the
reconstruction efficiency for combinatorial background was determined
from the data, as an adequate Monte Carlo model was not available.

\begin{figure}
\includegraphics[width=0.48\textwidth]{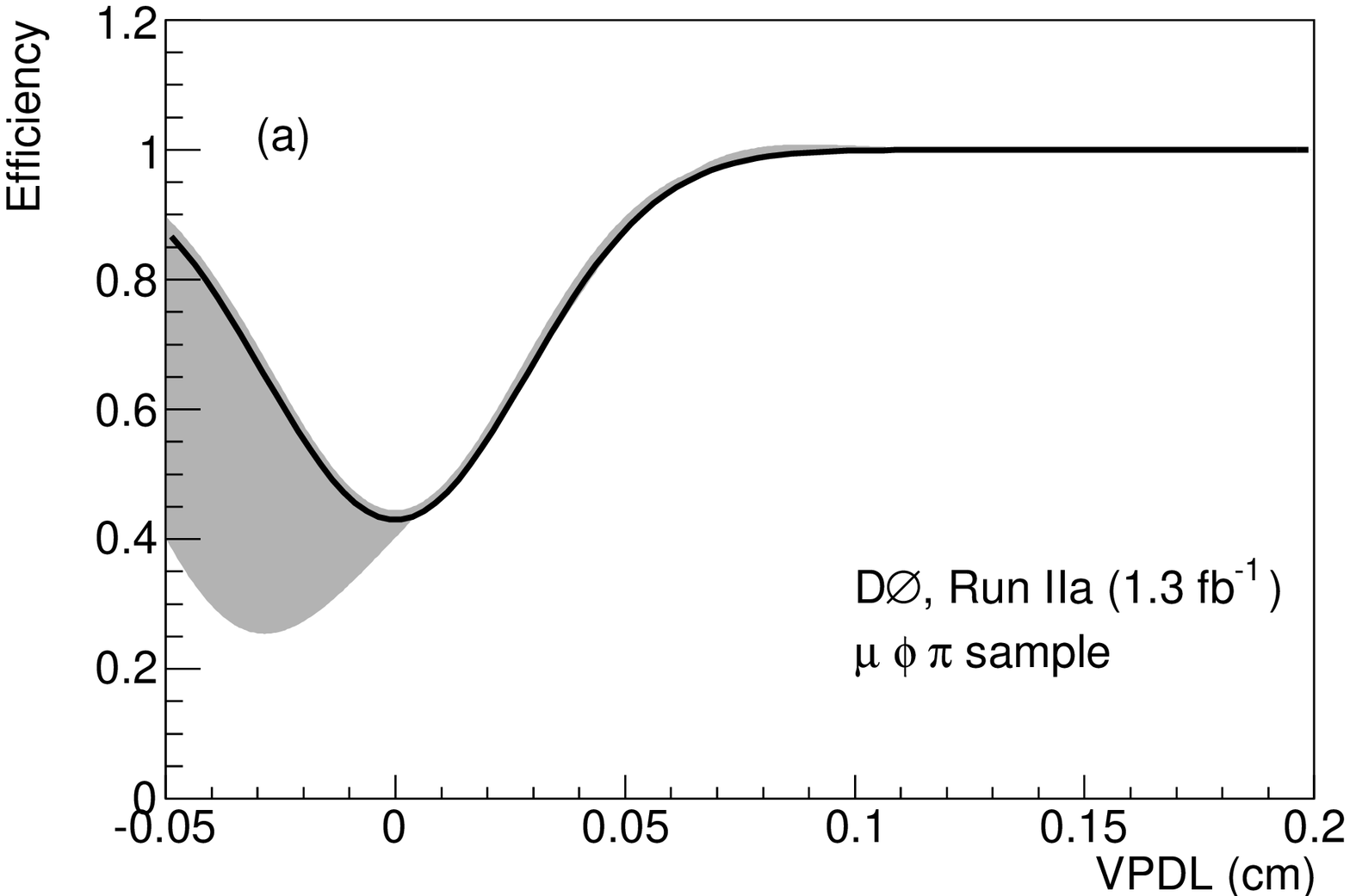}
\includegraphics[width=0.48\textwidth]{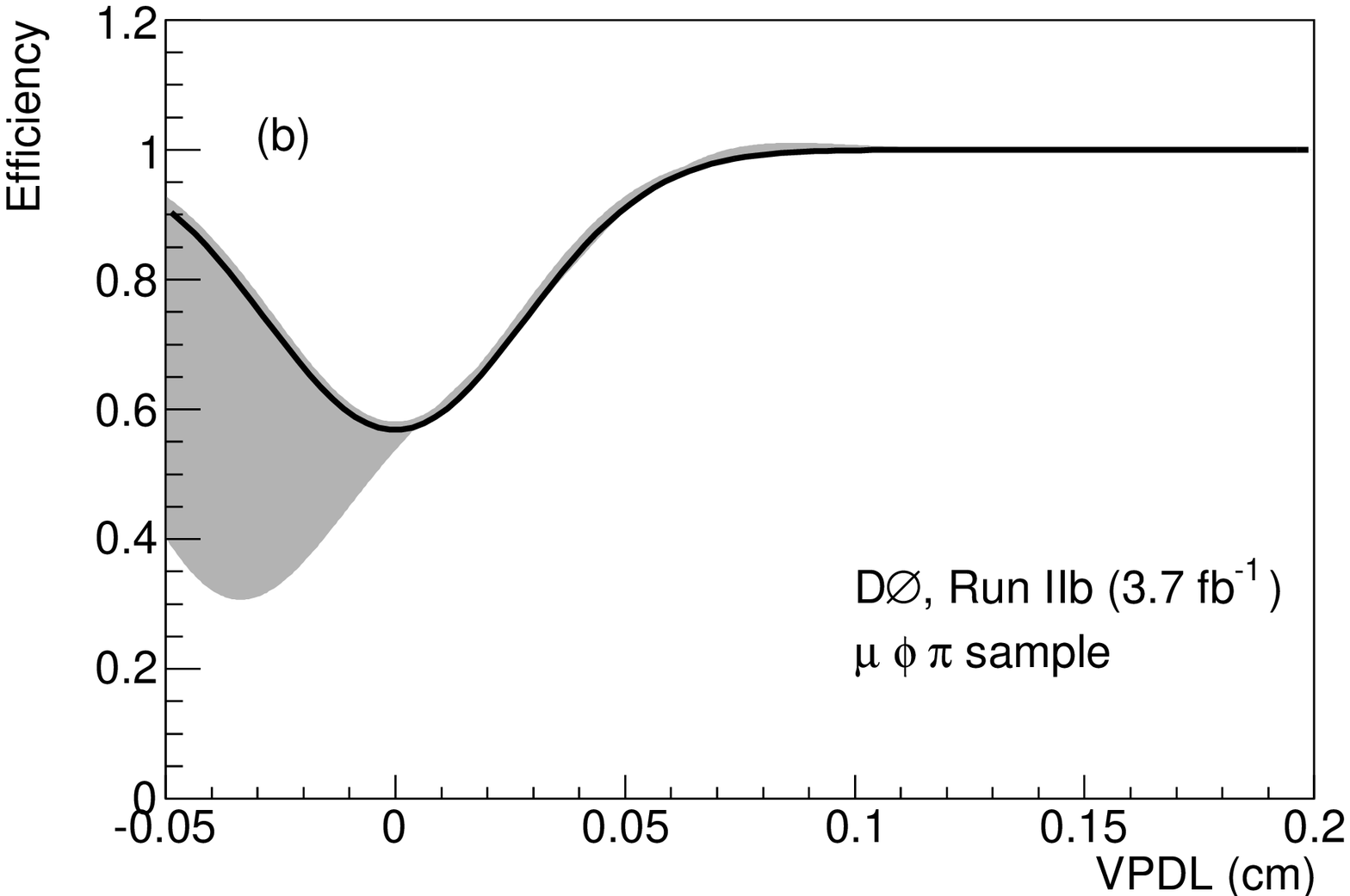}
\caption{\label{prd_fig5} The reconstruction efficiency as a function
  of VPDL for $\bstodsmux$ decays in the $\muphipi$ data samples.  
  Plot (a) is for Run~IIa and plot (b) is for Run~IIb. The width
  of the grey band indicates the range of the systematic uncertainty.}
\end{figure}

\begin{figure}
\includegraphics[width=0.48\textwidth]{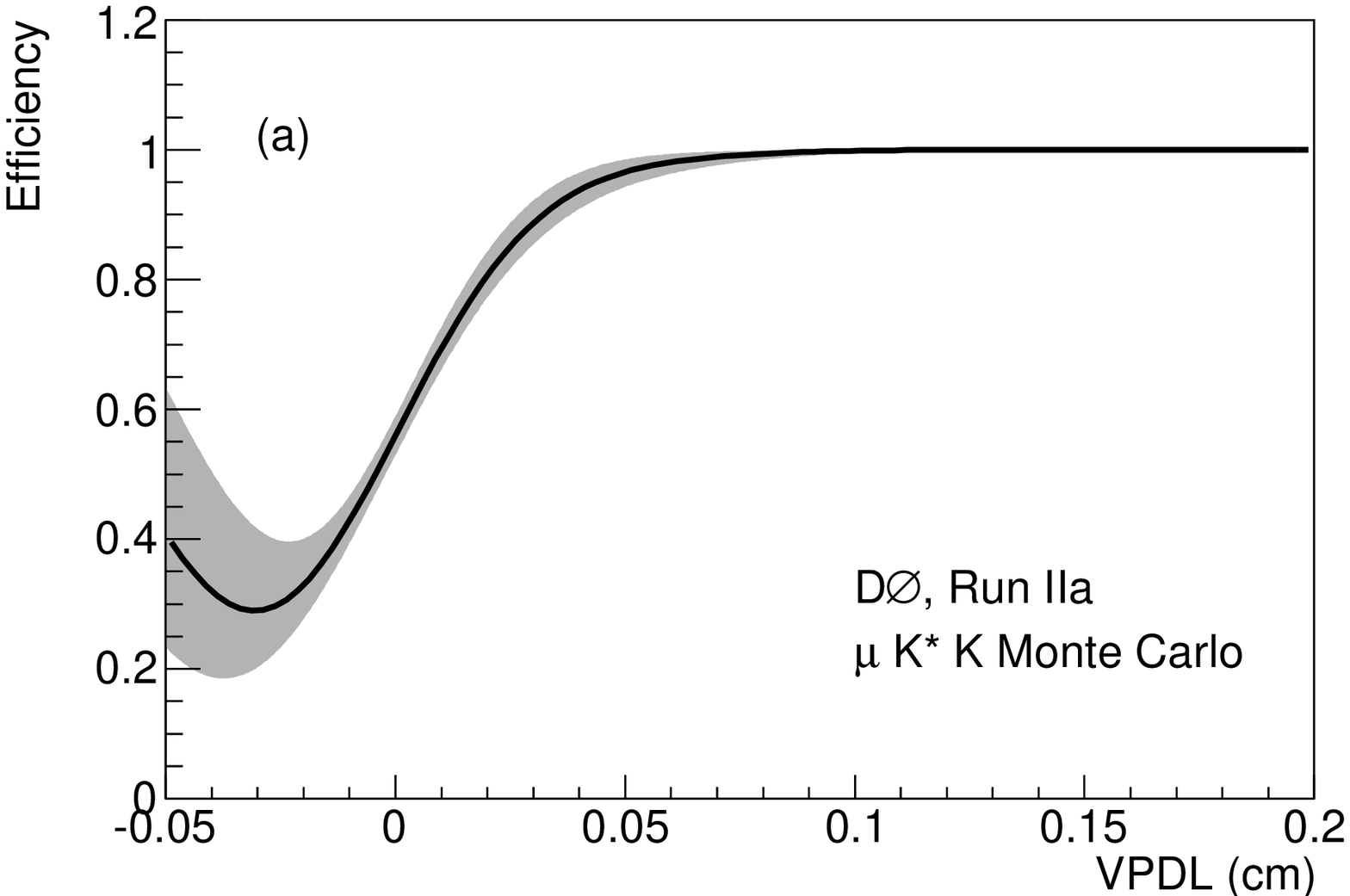}
\includegraphics[width=0.48\textwidth]{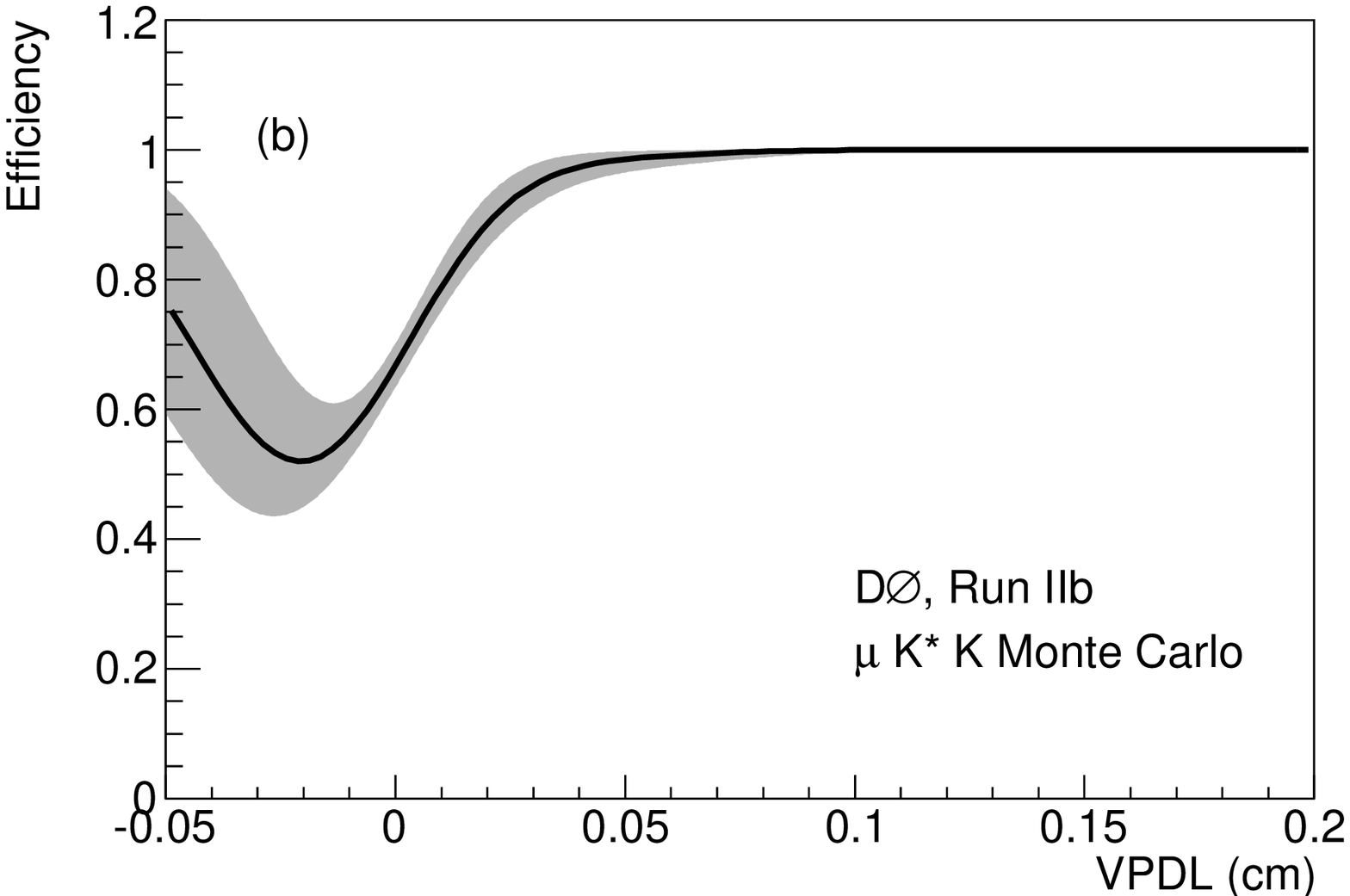}
\caption{\label{prd_fig6} The reconstruction efficiency as a function
  of VPDL for $\bstodsmux$ decays in the $\mukstk$ Monte Carlo samples.  Plot
  (a) is for Run~IIa and plot (b) is for Run~IIb. The width
  of the grey band indicates the range of the systematic uncertainty.}
\end{figure}

The decay-time dependent component of the signal efficiency was
obtained by considering the events passing all the selection criteria
except those dependent on the decay time.  This efficiency is shown as
a function of VPDL for the $\muphipi$ and the $\mukstk$ modes in
Figs.~\ref{prd_fig5} and \ref{prd_fig6}, respectively.  The VPDL range
extends below zero to account for the VPDL resolution. The efficiency
improvement visible in the low VPDL region in Run~IIb is due to the
addition of the Layer~0 detector.

The functional form that best describes the efficiency in the positive
VPDL region was chosen for the fit.  An analytic function is required
to properly normalize the $B_s^0$ decay time PDF. In the $\mukstk$
mode, where the signal efficiency was obtained from the simulation,
the VPDL value for the minimum was allowed to float in the fit.  In
the $\muphipi$ mode, where the signal efficiency was tuned to the
data, it was necessary to fix the minimum efficiency point to be at
VPDL~$=0$.  A systematic uncertainty accounting for this was obtained
by forcing a large negative variation in the efficiency in the
negative VPDL region, as indicated by the width of the bands in
Fig.~\ref{prd_fig5}.  In the $\mukstk$ mode, the systematic
uncertainty was obtained by simply varying the fit parameters within
their uncertainties, as indicated by the width of the bands in
Fig.~\ref{prd_fig6}.

The VPDL uncertainty $\sigma_l$ for a given event depends on the
uncertainties of the production and decay vertex positions, which in
turn depend on the track parameter uncertainties.  The procedure
described in Ref.~\cite{tune} was followed to tune the tracking
uncertainties using dijet and dimuon data samples.  Realistic
uncertainties were extracted from the track impact parameter pull
distributions, which were fitted by a single Gaussian function.  The
range for the fit was varied from between $\pm1\sigma$ and
$\pm3\sigma$.  Parametrizations of the dependence of these
uncertainties on the track momentum and polar angle were thus obtained
and applied to tune the track uncertainties in the analysis.  The
dimuon and dijet data samples yielded slightly different
results. Using the tuned track parameter uncertainties, the vertex
position uncertainty was also tuned, taking into account the tails,
using the pull distributions of the vertex positions of $J/\psi
\rightarrow \mu^+ \mu^-$ decays. Only the negative side of the pull
distributions was fitted, as the positive side includes a long-lived
component due to $J/\psi$ candidates from $B$ decays.  Separate vertex
uncertainty tunings were obtained for data before and after the
Layer~0 installation.  The tracking uncertainty differences observed
in the dimuon and dijet samples, combined with those associated with
the Gaussian fit range variation, yield a 5\% variation in the
resulting VPDL uncertainty. A 5\% systematic uncertainty on the VPDL
resolution was therefore included in the analysis.

\section{Sample Composition}

There are two contributions to the $D_s^-$ peaks in
Figs.~\ref{prd_fig3} and~\ref{prd_fig4}: the $B_{u,d,s} \to \mu^+
D_s^- X$ decays (including muons originating from $D_{(s)}$ and
$\tau^+$ decays) and the background occurring when the $D_s^-$ meson
originates from one $b$ or $c$ quark and the muon arises from the
decay of another quark. The fraction of the peak that is background
was determined from the data, as described later.  The contributions
from each $B$ decay channel, accounting for branching fractions, were
determined from the simulation.  Measured values for the branching
fractions were used~\cite{pdg}.  The exclusive branching fractions for
semileptonic $B_s^0$ decays to $D^{-}_s$, $D^{*-}_s$, ${D}_{s0}^{*-}$,
and ${D}_{s1}^{*-}$ have not been measured.  They were therefore
calculated from the measured branching fractions of the corresponding
decays for $B_d^0$ mesons, assuming the spectator model.  The
uncertainty associated with this assumption is expected to be
negligible~\cite{bigi} compared to the experimental branching fraction
uncertainties.

The contributions to the $\mu^+D_s^-$ signal prior to the application
of the decay-time dependent criteria in each mode are shown in
Table~\ref{tab:samplecomp}. The relative contribution of each source
varies with the invariant mass $M(\mu^+ D_s^-)$ of the reconstructed
$\mu^+ D_s^-$ system.  For example, an event with a high $M(\mu^+
D_s^-)$ is more likely to originate from a direct $B_s^0 \to \mu^+
\nu_{\mu}D_s^-$ decay, rather than from a $B_s^0 \to \mu^+
\nu_{\mu}D_s^{*-}$ decay, where the intermediate $D_s^{*-}$ decays to
$D_s^-X$. The relative contributions for each source were therefore
binned by invariant mass $M(\mu^+ D_s^-)$ for an improved model of the
sample composition.  

\begin{table}[htb]
\begin{center}
\caption{
Sample composition of the $\mu^+D_s^-$ signal. }
\label{tab:samplecomp}
\begin{ruledtabular}
\begin{tabular} {lrr}
\multicolumn{1}{c}{$\mu^+D_s^-$ Source} & \multicolumn{2}{c}{Fraction (\%)} \\
& \multicolumn{1}{c}{$\muphipi$} & \multicolumn{1}{c}{$\mukstk$} \\ \hline
$B_s^0 \to \mu^+ D_s^-\nu_{\mu}$  & 20.69 & 23.76 \\
$B_s^0 \to \mu^+ D_s^{*-}\nu_{\mu}$  & 63.26 & 60.22 \\
$B_s^0 \to \mu^+ D_{s0}^{*-}\nu_{\mu}$  & 1.56 & 1.65 \\
$B_s^0 \to \mu^+ D_{s1}^{*-}\nu_{\mu}$  & 3.23 & 3.16 \\
$B_s^0 \to \tau^+ D_s^-\nu_{\tau}$  & 1.05 & 0.25 \\
$B_s^0 \to D_s^-D_s^+X$  & 0.68 & 1.74 \\
$B_s^0 \to D_s^-DX$  & 0.68 & 0.30 \\
$B_s^0 \to D_s^+DX$  & 0.56 & 0.30 \\
$B^+ \to D_s^-DX$  & 3.57 & 2.94 \\
$B^0 \to D_s^-DX$  & 4.72 & 5.68 \\
\end{tabular}
\end{ruledtabular}
\end{center}
\end{table}

The mass PDF models the expected mass and width of the $K^+K^-\pi^{-}$
candidate for each source.  Four sources were considered in the
$\muphipi$ sample (see Fig.~\ref{prd_fig3}): the signal $\mu^+
D_s^-(\rightarrow \phi \pi^-)$; the accompanying mass peak due to
$\mu^+ D^{-} (\rightarrow \phi \pi^-)$; a small reflection (less than
1\% and therefore not visible in the figure) due to $\mu^+ D^{-}
(\rightarrow K^{+} \pi^{-} \pi^-)$, where the kaon mass is misassigned
to one of the pions; and combinatorial background.  Five sources were
considered in the $\mukstk$ sample (see Fig.~\ref{prd_fig4}): the
signal $\mu^+ D_s^-(\rightarrow K^{*0}K^-)$; the mass peak due to
$\mu^+ D^- (\rightarrow K^{*0}K^{-})$; a reflection due to $\mu^+ D^-
(\rightarrow K^{*0}\pi^-)$, where the pion is mistaken for a kaon; a
reflection due to $\mu^+ \Lambda_{c}^{-} (\rightarrow K^{*0}
\bar{p})$, where the antiproton is mistaken for a kaon; and
combinatorial background. The fractional contributions of these
sources were determined from the mass fits to the data.  The mass
distributions shown in Figs.~\ref{prd_fig3} and~\ref{prd_fig4} are
averages of the mass PDFs for each event.

The analysis of the $\mukstk$ mode is more challenging due to the
large $D^-$ reflection in the $D_s^-$ signal region (see
Fig.~\ref{prd_fig4}).  The $\mu^+D^-$ candidates arise from decays of all
three $B$ mesons, although the $B_s^0$ contribution is negligible.
The fraction of $\mu^+D^-$ candidates originating from $B_d^0$ decays
was assumed to be 80\% with 20\% arising from $B^+$
decays~\cite{evtgen}. The $B_d^0$ fraction can be measured
experimentally by exploiting the oscillating and non-oscillating
characteristics of the $B_d^0$ and $B^+$ mesons, respectively.  A
$B_d^0$ fraction of $0.93\pm0.04$ was thus determined from the
opposite-side tagged $\mukstk$ data sub-sample. The difference between
these two $B_d^0$ fractions was included in the $a_{fs}^s$ systematic
uncertainty as a possible bias.

\section{Likelihood Function}

All events with $1.72 < M(\phi\pi^-) < 2.22$~GeV$/c^2$ ($1.79 <
M(K^{*0}K^-) < 2.25$~GeV$/c^2$) were used in an unbinned fitting
procedure.  While the signal events in both modes are confined to a
narrower mass range, the events with masses outside the signal region
are needed to accurately describe the VPDL distribution for the
combinatorial background under the signal peaks. The total likelihood
$\mathcal{L}$ for $N$ selected events is the product of likelihoods
$L_j$ determined for each event $j$:
\begin{equation}
\mathcal{L}=\prod_{j=1}^{N}L_j,
\end{equation}
where
\begin{equation}
L_j=\sum_{i}\left[f_{i}P_{i}^{l}P_{i}^{\sigma_{l}}
  P_{i}^{y_{\mathrm{sel}}}P_{i}^{M(K^+K^-\pi^{-})}P_{i}^{d_{\mathrm{tag}}}\right].
\end{equation}
The sum is taken over products of the probability density functions
for different sources of $K^+K^-\pi^{-}$ candidates with fractions
$f_{i}$.  The distribution of the VPDL uncertainty is described by
$P_{i}^{\sigma_{l}}$.  The distribution of the likelihood ratio
selection variable is given by $P_{i}^{y_{\mathrm{sel}}}$. The mass
PDF is given by $P_{i}^{M(K^+K^-\pi^{-})}$. Finally,
$P_{i}^{d_{\mathrm{tag}}}$ is the distribution of the tagging variable
$d_{\mathrm{tag}}$. These PDFs were determined from data.

\begin{figure}
\includegraphics[width=0.48\textwidth]{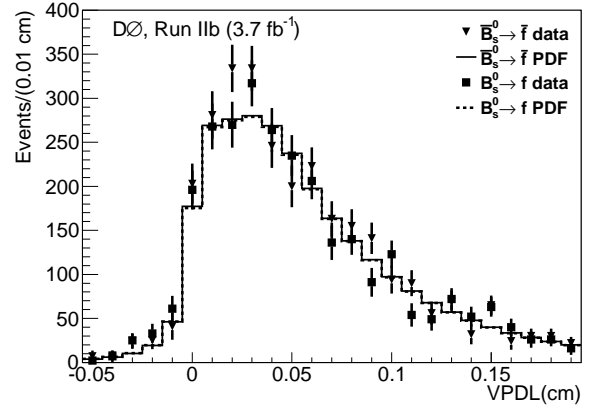}
\caption{\label{prd_fig7} The measured VPDL distribution
  for the Run~IIb $\muphipi$ subsample tagged as $B_{s}^{0}$
  ($\bar{B}_{s}^{0}$) in the initial state and $B_{s}^{0}$
  ($\bar{B}_{s}^{0}$) in the final state (i.e., {\em unmixed}). The
  PDFs for $\bar{B}^{0}_{s}(t)\to \bar{f}$ and $B^{0}_{s}(t)\to f$ are
overlaid.}
\end{figure}

\begin{figure}
\includegraphics[width=0.48\textwidth]{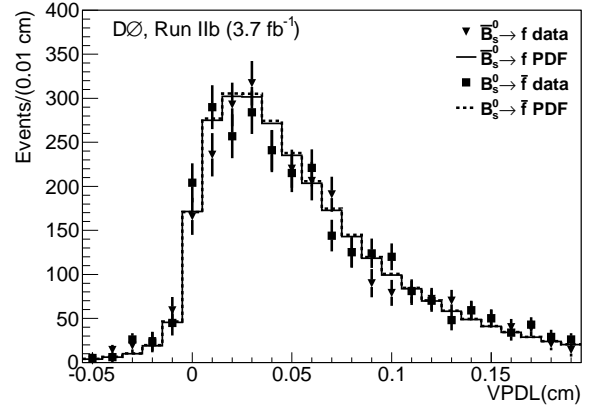}
\caption{\label{prd_fig8} The measured VPDL distribution for the
  Run~IIb $\muphipi$ subsample tagged as $B_{s}^{0}$
  ($\bar{B}_{s}^{0}$) in the initial state and $\bar{B}_{s}^{0}$
  ($B_{s}^{0}$) in the final state (i.e., {\em mixed}). The PDFs for
  $\bar{B}^{0}_{s}(t)\to f$ and $B^{0}_{s}(t)\to \bar{f}$ are
  overlaid.}
\end{figure}

The PDF $P_{i}^{l}$ for the measured VPDL $l$ was constructed as
follows.  The formulae for the decay rates of neutral $B_{s}^{0}$
mesons were taken from Ref.~\cite{tevb} assuming no direct CP
violation (i.e., $|\mathcal{A}_{f}|=|\bar{\mathcal{A}}_{\bar{f}}|$,
where $\mathcal{A}_{f}$ and $\bar{\mathcal{A}}_{\bar{f}}$ are the
decay amplitudes).  The asymmetry $a^{s}_{fs}$ modifies the decay
rates of mixed $B_{s}^{0}$ mesons as follows:
\begin{align}
\Gamma_{B^{0}_{s}(t)\to f} &= N_{f}|\mathcal{A}_{f}|^2e^{-\Gamma_{s} t}\notag\\
&[\cosh\left(\Delta\Gamma_{s} t/2\right)+\cos(\Delta m_{s} t) ]/2, \\
\Gamma_{B^{0}_{s}(t)\to \bar{f}} &= N_{f}|\bar{\mathcal{A}}_{\bar{f}}|^2(1-a^{s}_{fs})e^{-\Gamma_{s} t}\notag\\
&[\cosh\left(\Delta\Gamma_{s} t/2\right)-\cos(\Delta m_{s} t) ]/2,
\label{eq:gammaafs} \\
\Gamma_{\bar{B}^{0}_{s}(t)\to \bar{f}} &=
N_{f}|\bar{\mathcal{A}}_{\bar{f}}|^2e^{-\Gamma_{s} t}  \notag\\
&[\cosh\left(\Delta\Gamma_{s} t/2\right)+\cos(\Delta m_{s} t) ]/2, \label{eq:gammaafsbar} \\
\Gamma_{\bar{B}^{0}_{s}(t)\to f} &= N_{f}|\mathcal{A}_{f}|^2(1+a^{s}_{fs})e^{-\Gamma_{s} t}\notag\\
&[\cosh\left(\Delta\Gamma_{s} t/2\right)-\cos(\Delta m_{s} t) ]/2,
\end{align}
where $N_{f}$ is the normalization to the total number of $B^{0}_{s}$
mesons. Using the relationship $c t(B^0_s) = l \mathcal{K}$ discussed
previously, these decay-rate PDFs can be written in terms of VPDL. All
reconstructed events were divided into four samples corresponding to
the final- and initial-state tags.  The dilution of the initial-state
tagging leads to a mixture of $B_{s}^{0}$ and $\bar{B}_{s}^{0}$
initial states in these samples, e.g., the sample tagged as
$B_{s}^{0}$ in the initial state and $\bar{B}_{s}^{0}$ in the final
state has the PDF 
\begin{equation}
P_{B_{s}^{0}\bar{B}_{s}^{0}}^{l}=
\Gamma_{B^{0}_{s}(t)\to\bar{f}}{\sl p}_{\mathrm{cor}}
+ \Gamma_{\bar{B^{0}_{s}}(t)\to \bar{f}}
(1-{\sl p}_{\mathrm{cor}}),
\end{equation}
where $\Gamma_{B^{0}_{s}(t)\to\bar{f}}$ is the PDF in
Eq.~\ref{eq:gammaafs} and $\Gamma_{\bar{B}^{0}_{s}(t)\to \bar{f}}$ is
the PDF in Eq.~\ref{eq:gammaafsbar}. As an example, the average VPDL
PDFs for {\em unmixed} events (i.e., $P_{B_s^0B_s^0}^l$ and
$P_{\bar{B}_s^0\bar{B}_s^0}^l$) and {\em mixed} events (i.e.,
$P_{B_s^0\bar{B}_s^0}^l$ and $P_{\bar{B}_s^0B_s^0}^l$) for the Run~IIb
$\muphipi$ sample are shown in Figs.~\ref{prd_fig7}
and~\ref{prd_fig8}, respectively. For the figures, the combinatorial
background was subtracted using sidebands of the $K^+K^-\pi^-$
invariant mass distributions. A contribution describing the background
due to fake vertices around the primary vertex was included into the
PDF. The slight differences in the PDFs for the mixed events in
Fig.~\ref{prd_fig8} give rise to a non-zero asymmetry $a_{fs}^s$.

The decay-rate PDF for the semileptonic decays $B_{d}^{0}\to
\mu^{+}D^{-}X$ (the events comprising the peaks at 1.87~GeV/$c^2$ in
Figs.~\ref{prd_fig3} and~\ref{prd_fig4}) is the same as for the
$B_{s}^{0}$ decays with the corresponding parameters changed
accordingly. In particular, the $B_{d}^{0}$ semileptonic asymmetry
$a_{fs}^{d}$ was introduced and determined from the fit.

\section{Detector Asymmetries}

The PDFs are modified to account for the detector charge
asymmetries~\cite{d0dimu,d0bsunt}:
\begin{align}
 A_{\mu} = 
&(1+\gamma A_{\mathrm{det}})
(1+q\beta\gamma A_{\mathrm{ro}})
(1+q\gamma A_{\mathrm{fb}})\cdot \notag\\
&(1+\beta\gamma A_{\beta\gamma})(1+q\beta A_{q\beta}), 
\end{align}
where $\beta$ is the toroid polarity, $\gamma$ is the sign of the muon
pseudorapidity, and $q$ is the muon charge.  $A_{\mathrm{det}}$
accounts for any asymmetry due to differences in the north ($z<0$) and
south ($z>0$) ends of the detector (the proton beam travels from north
to south).  The range-out asymmetry $A_{\mathrm{ro}}$ reflects the
difference in acceptance of muons that bend towards as opposed to away
from the beam line (for details, see Ref.~\cite{d0dimu}).  The
forward-backward asymmetry $A_{\mathrm{fb}}$ reflects the fact that
positively charged muons tend to go in the direction of the proton
beam whereas negatively charged muons typically go in the antiproton
beam direction.  $A_{\beta\gamma}$ is a second-order correction to
$A_{\mathrm{ro}}$ that is non-zero only if $a^s_{fs}$ and
$A_{\mathrm{ro}}$ are both non-zero.  $A_{q\beta}$ is a detector
asymmetry between tracks bending towards $\eta<0$ and tracks bending
towards $\eta>0$.  The likelihood for each event was multiplied by the
detector charge asymmetry corrections for the muon from the
$B_{s}^{0}$ semileptonic decay and for the muon used for the OST when
it is present.  Both kaons in the $\muphipi$ signal sample originate
from the $\phi$ decay and therefore have the same transverse momentum
threshold.  Since this is not the case in the $\mukstk$ sample, the
momentum dependent kaon reconstruction asymmetry~\cite{jpsik} was
taken into account.

\section{Background Description}

The background where the $D_s^-$ meson and the muon do not come from
the same parent particle gives rise to fake vertices and a VPDL
distribution that peaks around zero.  Its shape was modeled by two
Gaussian functions, and its contribution was estimated from decay-time
fits to be approximately 8\% for the total sample and 3\% for the
opposite-side tagged sub-sample for the $\muphipi$ mode.  The
corresponding contributions are 4\% and 1\% for the $\mukstk$
mode. The fake vertex background fraction is lower in the $\mukstk$
mode due to the higher kinematic selection criteria.  Several
contributions to the combinatorial background that have different VPDL
distributions were considered.  True prompt background was modeled
with a Gaussian function. The long-lived combinatorial background is
dominated by misreconstructed heavy flavor decays. This background was
modeled with an exponential function convoluted with the VPDL
resolution, including a component ($\approx 60\%$ of all long-lived
background contributions) oscillating with a frequency of $\Delta
m_d$.  An asymmetry parameter $a_{bg}$ was introduced into the PDFs
for this component by analogy with Eqs.~\ref{eq:gammaafs} and
\ref{eq:gammaafsbar}.  This parameter absorbs possible asymmetries in
the combinatorial background that were unaccounted for.  The unbinned
likelihood fit of the total sample was used to determine the various
fractions of signal and backgrounds and the background VPDL
parametrizations.

\section{Results}

The $B_{s}^{0}$ oscillation frequency $\Delta
m_{s}=17.77\pm0.12$~ps$^{-1}$~\cite{cdfdms} was fixed in the fit. In
the $\muphipi$ mode, the width difference was fixed to
$\Delta\Gamma_{s}=0.09\pm0.05$~ps$^{-1}$~\cite{pdg}.  In the $\mukstk$
mode, the ratio $\Delta\Gamma_{s}/\Gamma_s=0.069^{+0.058}_{-0.062}$
was fixed in the fit.  First, the $\muphipi$ and $\mukstk$ samples
were fitted separately (see Table~\ref{tab:sigdms}). The results for the
charge asymmetries are consistent between these samples.
\begin{table}[htb]
\begin{center}
\caption{
Asymmetries with statistical uncertainties.}
\label{tab:sigdms}
\begin{ruledtabular}
\begin{tabular}{cD{,}{\pm}{5}D{,}{\pm}{5}D{,}{\pm}{5}}
                 & \multicolumn{1}{c}{$\muphipi$} & \multicolumn{1}{c}{$\mukstk$} & \multicolumn{1}{c}{Combined} \\ \hline
$a_{fs}^{s}\times 10^{3}$ &    -7.0,9.9  & 20.3,24.9  & -1.7,9.1          \\ 
$a_{fs}^{d}\times 10^{3}$ &    -21.4, 36.3  & 50.1, 19.5  & 40.5, 16.5        \\
$a_{bg}\times 10^{3}$ &        -2.2 , 10.6  & -0.1 , 13.5  & -3.1, 8.3       \\
$A_{\mathrm{fb}}\times 10^{3}$ &        -1.8 , 1.5  & -2.0 , 1.5  & -1.9, 1.1        \\
$A_{\mathrm{det}}\times 10^{3}$ &       3.2 , 1.5  & 3.1 , 1.5  & 3.1, 1.1         \\ 
$A_{\mathrm{ro}}\times 10^{3}$ &        -36.7 , 1.5  & -30.2 , 1.5  & -33.3, 1.1       \\
$A_{\beta\gamma}\times 10^{3}$&1.1 , 1.5  & 0.2 , 1.5  & 0.6, 1.1     \\ 
$A_{q\beta}\times 10^{3}$ &    4.3 , 1.5  & 2.0 , 1.5  & 3.1, 1.1     \\ 
\end{tabular}
\end{ruledtabular}
\end{center}
\end{table}
The results of a fit to the combined $\muphipi$ and $\mukstk$ samples
are shown in the last column of Table~\ref{tab:sigdms}.  The measured
value for the flavor-specific asymmetry is $a_{fs}^{s}=[-1.7\pm
9.1]\times 10^{-3}$, where the uncertainty is statistical. The
detector charge asymmetry $A_{\mathrm{ro}}$ is observed to be
non-zero, as expected~\cite{d0dimu}.  The other detector asymmetries
are consistent with zero, as is the physics asymmetry $a_{bg}$.  There
is no correlation between the detector charge asymmetries, either
amongst themselves or with the physics asymmetries.  There is a
$5.2\%$ ($13.6\%$) correlation between $a_{fs}^{s}$ and $a_{fs}^{d}$
($a_{bg}$). The non-zero value of $a_{fs}^{d}$ is discussed in the
next section.

\begin{table}[h]
\begin{center}
\caption{
Systematic uncertainties.}
\label{tab:sys}
\begin{ruledtabular}
\begin{tabular}{lr}
                 & \multicolumn{1}{c}{$\sigma(a_{fs}^{s})\times 10^{3}$} \\ \hline
Kaon asymmetry set to 0 &  -1.24          \\ 
Kaon asymmetry scaled by 2 &  1.30          \\ 
Signal fraction $-1\sigma$ & -0.76          \\ 
Signal fraction $+1\sigma$ & 0.47             \\ 
Dilution scaled by 0.9 & -0.19        \\ 
Dilution scaled by 1.1 & 0.21        \\ 
$\mu$ trigger efficiency low & -0.03             \\ 
$\mu$ trigger efficiency high &  0.00          \\ 
Decay-time dependent efficiency low & 0.15           \\ 
Decay-time dependent efficiency high & -0.01          \\ 
VPDL resolution scaled by 0.95 &  0.03          \\ 
VPDL resolution scaled by 1.05 &  -0.03          \\ 
BF $\bs\to D_s^- D_s^+$ &  0.00          \\ 
BF $\bs\to\mu^+D_s^{(*)-}X$ & -0.10          \\ 
Relative BF $\bs\to\mu^+\nu_{\mu}D_s^-$ low &  0.01          \\ 
Relative BF $\bs\to\mu^+\nu_{\mu}D_s^-$ high & -0.05           \\ 
$B_d^0$ fraction in $\mu^+D^-$ candidates set to 93\% & -0.24 \\ 
Fake vertex background low  & -0.13          \\ 
Fake vertex background high & -0.04         \\ 
Prompt combinatorial background low  &  0.01          \\ 
Prompt combinatorial background high &  -0.01          \\ 
$\Delta\Gamma_s$ $-1\sigma$ & 0.00            \\ 
$\Delta\Gamma_s$ $+1\sigma$ & -0.01         \\ 
$\Delta m_{s}$ $-1\sigma$ & -0.01         \\ 
$\Delta m_{s}$ $+1\sigma$ & 0.02          \\ 
\hline
Total & $^{+1.41}_{-1.50}$\\
\end{tabular}
\end{ruledtabular}
\end{center}
\end{table}

\section{Systematic Uncertainties}

The sources contributing to the total $a^{s}_{fs}$ systematic
uncertainty of $^{+1.4}_{-1.5}\times10^{-3}$ are listed in
Table~\ref{tab:sys}.  The largest contribution is due to the
momentum-dependent kaon asymmetry.  The fitted value of $a_{fs}^{d}$
is sensitive to this asymmetry due to the large $D^-$ reflection under
the $D_s^-$ signal peak in the $\mukstk$ mode.  To account for this,
and for the correlation between $a_{fs}^d$ and $a_{fs}^s$, the kaon
asymmetry was varied between zero and twice its nominal value and the
likelihood fit was repeated in each case. Setting the kaon asymmetry
to zero yielded a fitted value of
$a^{d}_{fs}=[-0.2\pm16.5]\times10^{-3}$, whereas scaling it by two
yielded $a^{d}_{fs}=[81.2\pm16.6]\times10^{-3}$. The resulting change
in the fitted $a^{s}_{fs}$ value is taken as the associated systematic
uncertainty, given in Table~\ref{tab:sys}.

The same procedure was followed for the remaining sources of
uncertainty. The signal fraction, that is, the number $D_s^-$ and
$D^-$ signal events divided by the total number of events, obtained
from the mass fits was varied up and down by $1\sigma$.  The dilution
of the opposite-side tagging algorithm was scaled by 0.9 and 1.1.  The
muon trigger efficiency was varied within its experimental
uncertainties.  The decay-time dependent reconstruction efficiency as
a function of VPDL was varied, as previously described.  The VPDL
resolution was scaled by 0.95 and 1.05 to account for the tracking
uncertainty differences previously discussed.  The fake vertex
background, where the $D_s^-$ meson and the muon do not come from the
same parent particle, was varied within its experimental range as was
the prompt component of the combinatorial background.  The
$B_s^0-\bar{B}_s^0$ width difference $\Delta \Gamma_{s}$ and the
oscillation frequency $\Delta m_{s}$ were varied within their
experimental uncertainties. The relative branching fractions (BF) for
the exclusive semileptonic $B_s^0$ decays were varied within their
predicted uncertainties in such a way as to keep the total inclusive
fraction constant.  To model a reduced signal fraction, the $\bs\to
D_s^- D_s^+$ branching fraction was increased by $1\sigma$ and the
$\bs\to \mu^+ D_s^{(*)-} X$ inclusive branching fraction was decreased
by $1\sigma$. To account for a possible bias associated with the
fraction of $\mu^+D^-$ candidates in the $\mu^+K^{*0}K^-$ sample
originating from $B_d^0$ decays (see Fig.~\ref{prd_fig4}), this
fraction was changed from 0.80 to 0.93 in the likelihood fit.  The total
systematic uncertainty was determined by adding all the signed
contributions in quadrature.

\section{Conclusion}

In summary, using $\bstodsmux$ decays with $\dstophipi$, $\phitokk$
and $\dstokstk$, $\ksttokpi$, in combination with a decay-time
analysis including initial-state flavor tagging, we measure the
asymmetry in mixed semileptonic $B_{s}^{0}$ decays to be
$a_{fs}^{s}=[-1.7\pm 9.1\mathrm{(stat)}^{+1.4}_{-1.5}\mathrm{(syst)}]
\times 10^{-3}$.  This measurement supercedes the D0 time-integrated 
analysis of semileptonic $B_s^0$ decays~\cite{d0bsunt}, which yielded
$a^s_{fs}=[24.5\pm 19.3\mathrm{(stat)} \pm 3.5\mathrm{(syst)}]\times
10^{-3}$.  Our result is also consistent with the value
$a^s_{fs}=(-6.4\pm 10.1)\times 10^{-3}$ extracted~\cite{d0comb} from
the D0 time-integrated analysis of inclusive same-sign dimuon
events~\cite{d0dimu}.  While the present result is the most precise
measurement of the semileptonic $B_{s}^{0}$ asymmetry, improved
precision is needed to establish evidence of CP violation due to new
physics in $B_s^0$ mixing.

%
We thank the staffs at Fermilab and collaborating institutions, 
and acknowledge support from the 
DOE and NSF (USA);
CEA and CNRS/IN2P3 (France);
FASI, Rosatom and RFBR (Russia);
CNPq, FAPERJ, FAPESP and FUNDUNESP (Brazil);
DAE and DST (India);
Colciencias (Colombia);
CONACyT (Mexico);
KRF and KOSEF (Korea);
CONICET and UBACyT (Argentina);
FOM (The Netherlands);
STFC and the Royal Society (United Kingdom);
MSMT and GACR (Czech Republic);
CRC Program, CFI, NSERC and WestGrid Project (Canada);
BMBF and DFG (Germany);
SFI (Ireland);
The Swedish Research Council (Sweden);
CAS and CNSF (China);
and the
Alexander von Humboldt Foundation (Germany).


\begin{thebibliography}{99}
%
\bibitem[a]{alton}
Visitor from Augustana College, Sioux Falls, SD, USA.
\bibitem[b]{askew,atramentov,gershtein}
Visitor from Rutgers University, Piscataway, NJ, USA.
\bibitem[c]{burdin}
Visitor from The University of Liverpool, Liverpool, UK.
\bibitem[d]{luna-garcia}
Visitor from Centro de Investigacion en Computacion - IPN,
  Mexico City, Mexico.
\bibitem[e]{podesta-lerma}
Visitor from ECFM, Universidad Autonoma de Sinaloa, Culiac\'an, Mexico.
\bibitem[f]{voutilainen}
Visitor from Helsinki Institute of Physics, Helsinki, Finland.
\bibitem[g]{weber}
Visitor from Universit{\"a}t Bern, Bern, Switzerland.
\bibitem[h]{wenger}
Visitor from Universit{\"a}t Z{\"u}rich, Z{\"u}rich, Switzerland.
\bibitem[\ddag]{deceased}
Deceased.

%
\vskip 0.25cm
%
\bibitem{lenz_nierste} A.~Lenz and U.~Nierste, JHEP {\bf 0706}, 072 (2007).
\bibitem{hou} W.-S.~Hou, M.~Nagashima and A.~Soddu, Phys. Rev. D {\bf
  76}, 016004 (2007).
\bibitem{cdfjpsiphi} T.~Aaltonen {\it et al.} (CDF Collaboration), Phys. Rev. Lett. {\bf 100}, 161802 (2008).
\bibitem{d0jpsiphi} V.~M.~Abazov {\it et al.} (D0 Collaboration), Phys. Rev. Lett. {\bf 101}, 241801 (2008).
\bibitem{hfag} Heavy Flavour Averaging Group (HFAG), ``Results for the PDG 2009 web update'',  
{\tt http://www.slac.stanford.edu/xorg/hfag/}.
\bibitem{private} This $a^s_{fs}$ value is calculated by the authors 
and neglects correlations between systematic uncertainties.
\bibitem{lumi}T.~Andeen {\it et al.}, FERMILAB-TM-2365 (2007).
\bibitem{d0dimu} V.~M.~Abazov {\it et al.} (D0 Collaboration), Phys. Rev. D {\bf 74}, 092001 (2006).
\bibitem{D0} V.~M.~Abazov {\it et al.} (D0 Collaboration), Nucl. Instrum. Methods Phys. Res. A
{\bf 565}, 463 (2006).
\bibitem{l0}
D.~Tsybychev {\it et al.} (D0 Collaboration),
Nucl. Instrum. Methods Phys. Res. A {\bf 582}, 701 (2007).
\bibitem{NIM} V.~M.~Abazov {\it et al.}, Nucl. Instrum. Meth. A {\bf
552}, 372 (2005).
\bibitem{durham} S.~Catani {\em et al.}, Phys.\ Lett.\ B {\bf 269}, 432
  (1991). Durham jets with the $p_T$ cut-off parameter set at 15~GeV/$c$.
\bibitem{d0bsmix} V.~M.~Abazov {\it et al.} (D0 Collaboration), Phys. Rev. Lett. {\bf 97}, 021802 (2006).
\bibitem{thesis} S.~Beale, Ph.D. Dissertation, York University,
  Toronto, Canada (2010), FERMILAB-THESIS-2010-06.
\bibitem{vertex} 
  J.~Abdallah {\it et al.}  (DELPHI Collaboration),
  Eur.\ Phys.\ J.\ C {\bf 32}, 185 (2004).
\bibitem{like_ratio}
G.~Borisov,
  Nucl. Instrum. Methods Phys. Res. A {\bf 417}, 384 (1998).
\bibitem{ost} V.~M.~Abazov {\it et al.} (D0 Collaboration), Phys. Rev. D {\bf 74}, 112002 (2006).
\bibitem{pdg} C.~Amsler {\it et al.}, Phys.\  Lett.\  B {\bf 667}, 1 (2008).
\bibitem{pythia}
T. Sj\"{o}strand {\it et al.}, Comput. Phys. Commun. {\bf 135}, 238
(2001).
\bibitem{evtgen} D.~J.~Lange, Nucl. Instrum. Methods Phys. Res. A {\bf 462}, 152 (2001); 
for details see {\tt http://www.slac.stanford.edu/\~\!lange/EvtGen}.
\bibitem{geant} 
 R. Brun and F. Carminati, 
CERN Program Library Long Writeup W5013 (unpublished).
\bibitem{tune}
G.~Borisov and C.~Mariotti,
Nucl.\ Instrum.\ Methods Phys. Res. A {\bf 372}, 181 (1996).
\bibitem{bigi} I.~Bigi, M.~Shifman and N.~Uraltsev,
  Ann. Rev. Nucl. Part. Sci. {\bf 47}, 591 (1997).
\bibitem{tevb} K.~Anikeev {\em et al.}, arXiv:hep-ph/0201071 (2002); see Eqs.~1.78 and 1.79.
\bibitem{d0bsunt} V.~M.~Abazov {\it et al.} (D0 Collaboration), Phys. Rev. Lett. {\bf 98}, 151801 (2007).
\bibitem{jpsik}  V.~M.~Abazov  {\em et al.} (D0 Collaboration), Phys.\ Rev.\ Lett.\ {\bf 100}, 211802 (2008).
\bibitem{cdfdms} A.~Abulencia {\em et al.} (CDF Collaboration), Phys.\ Rev.\ Lett.\  {\bf 97}, 242003 (2006). 
\bibitem{d0comb} V.~M.~Abazov {\it et al.} (D0 Collaboration), Phys. Rev. D {\bf 76}, 057101 (2007).
\end{thebibliography}
\end{document}